\documentclass{INTERSPEECH2023}
\usepackage{amsmath,graphicx}
\usepackage[ruled]{algorithm2e}

\usepackage{boldline}
\usepackage{array, makecell}
\usepackage{multicol}
\usepackage{multirow}
\usepackage{color}
\usepackage{booktabs}
\usepackage{tabularx,adjustbox}
\usepackage{subfigure}
\usepackage{setspace}

\newcommand{\PreserveBackslash}[1]{\let\temp=\\#1\let\\=\temp}
\newcolumntype{C}[1]{>{\PreserveBackslash\centering}p{#1}}
\newcolumntype{R}[1]{>{\PreserveBackslash\raggedleft}p{#1}}
\newcolumntype{L}[1]{>{\PreserveBackslash\raggedright}p{#1}}
\newcommand\thefont{\expandafter\string\the\font}

\interspeechcameraready 

\title{Feature Normalization for Fine-tuning Self-Supervised Models \\ in Speech Enhancement}
\name{Hejung Yang, Hong-Goo Kang}
\address{Dept. of Electrical and Electronic Engineering, Yonsei University, South Korea}
\email{hejung.yang@dsp.yonsei.ac.kr, hgkang@yonsei.ac.kr}

\begin{document}

\maketitle

 \setlength{\tabcolsep}{15pt}
\begin{abstract}
Large, pre-trained representation models trained using self-supervised learning have gained popularity in various fields of machine learning because they are able to extract high-quality salient features from input data.
As such, they have been frequently used as base networks for various pattern classification tasks such as speech recognition.
However, not much research has been conducted on applying these types of models to the field of speech signal generation.
In this paper, we investigate the feasibility of using pre-trained speech representation models for a downstream speech enhancement task.
To alleviate mismatches between the input features of the pre-trained model and the target enhancement model, we adopt a novel feature normalization technique to smoothly link these modules together.
Our proposed method enables significant improvements in speech quality compared to baselines when combined with various types of pre-trained speech models.
%More generally, it can be applied to any type of downstream task that utilizes such pre-trained models.
\end{abstract}

\section{Introduction}
\label{sec:intro}

Large, pre-trained models trained using self-supervised learning have become popular in various speech processing tasks \cite{oord2018representation, baevski2020wav2vec}. Due to their ability to leverage representations learned from large amounts of unlabeled data, these models have been successfully applied to a wide variety of downstream tasks such as automatic speech recognition (ASR), speaker verification (SV) \cite{xia2021self}, and audio scene classification \cite{tripathi2021self}.

Depending on the characteristics of their training objectives, these pre-trained models can be categorized into either generative models or contrastive models \cite{liu2021self}. In generative models, decoders are trained to reconstruct masked frames or future frames; they include models such as APC \cite{chung19unsupervised}, Mockingjay \cite{liu2020mockingjay}, and TERA \cite{liu2021tera}. Contrastive models are trained by utilizing similarities and differences between latent embeddings. A representative example is wav2vec 2.0 \cite{baevski2020wav2vec}, and several variants such as HuBERT \cite{hsu2021hubert} and WavLM \cite{chen2022wavlm} have been proposed that adopt various types of regularizers for each target objective. Meanwhile, some models utilize both generative and contrastive objectives, such as PASE+ \cite{ravanelli2020multi}.
Most upstream models adopting self-supervised pre-training for speech applications have focused on solving discriminative tasks \cite{yang2021superb}. Recently, several attempts \cite{huang2022investigating, hung2022boosting} have been made to apply these models for speech enhancement (SE) or speech separation tasks. 

\begin{figure}[t]
  \centering
  \includegraphics[width=\linewidth]{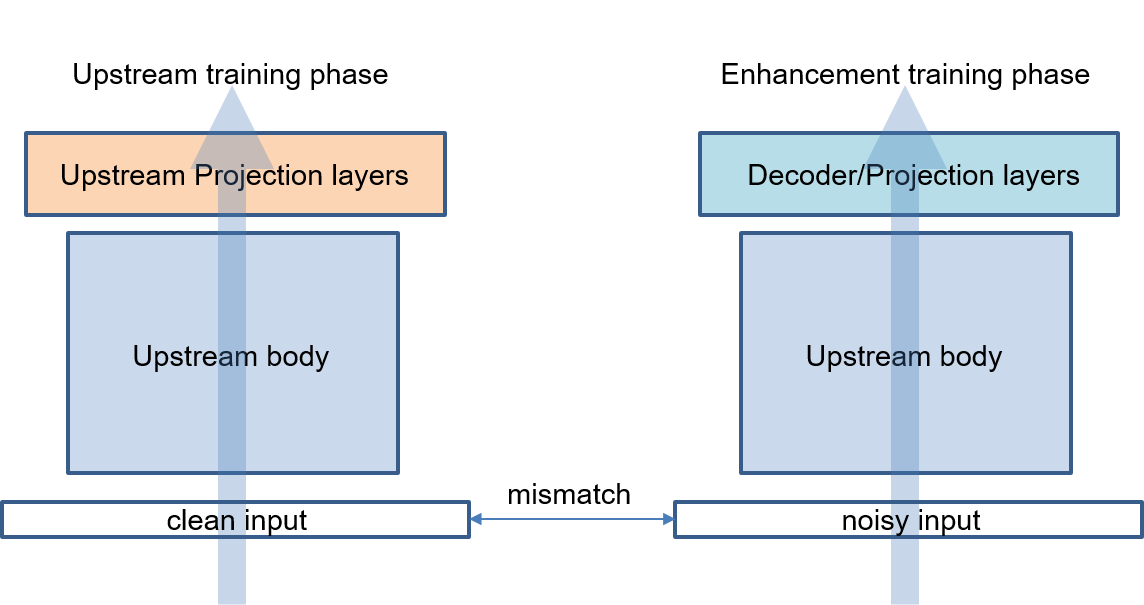}
  \caption{Domain mismatch when using pre-trained upstream body trained on clean data for speech enhancement task.}
  \label{fig:1}
  \vspace{-8pt}
\end{figure}

When pre-trained speech models are used for SE, it is inevitable that they encounter a domain mismatch problem. This is because most of these pre-trained models are trained on clean data, while the SE task fundamentally requires corrupted or noisy speech inputs (Fig. \ref{fig:1}). Although this domain mismatch problem can be relieved by pre-training the upstream models with both noisy and clean data \cite{hsu2021robust}, this requires a huge amount of data and additional training time, which takes much longer than fine-tuning a downstream model with labeled data.
Another drawback is that this precludes the use of well-trained upstream models that have been made publicly available online.

In this paper, we propose an effective feature normalization technique that facilitates the use of representations from pre-trained speech models on a downstream SE task.
To alleviate the dissonance between the noisy inputs fed into the downstream model in the fine-tuning phase and the initial weight distribution of the upstream body trained with clean speech, we normalize the latent features from the noisy input with its clean referential statistics, which can be estimated by feeding clean targets to the frozen upstream model. 
By adjusting the degree of normalization during training, the model can smoothly change its input domain from clean to noisy, thereby improving SE performance in the end.
More generally, it can be applied to any type of downstream task that utilizes such pre-trained models.

Our contributions are as follows: 1) With our proposed feature normalization technique applied during training, we show that it is possible to achieve better SE performance when using the representations from pre-trained speech models without introducing any additional parameters or training losses. 2) We can directly utilize these pre-trained models without the need for any additional domain-adaptation to calibrate the pre-trained models or to train them from scratch.

The rest of the paper is organized as follows. After describing the related works in section \ref{sec:related}, base downstream models trained on top of different upstream architectures are introduced in section \ref{sec:base}. Section \ref{sec:featnorm} depicts feature normalization algorithm, and results are presented in section \ref{sec:setup} and \ref{sec:results}.

\section{Related work}
\label{sec:related}

Previous works have shown that the domain mismatch between the pre-training and fine-tuning phases degrades the performance of the downstream models that utilize pre-trained speech representations \cite{hsu2021robust, meng2022don}. Utilizing target-domain data in the pre-training stage can mitigate the degradation \cite{hsu2021robust, hwang2022large}, but this requires additional training which removes the efficiency we can obtain by using existing pre-trained models.

One algorithm for domain adaptation is Domain Adversarial Training (DAT), where the domain classifier is trained adversarially so that the representation cannot be identified by the domain context \cite{ganin2016domain}. DAT has been applied to various supervised tasks including domain-specific ASR, SV and SE \cite{sun2018domain, wang2018unsupervised, liao2018noise}. However, it requires multi-task learning for training the domain classifier with a categorization of domain classes.
Other methods include residual adapters \cite{fan2022draft} or auxiliary contrastive loss \cite{chen2021self}. However, injection of new parameters dedicated for domain adaptation or training paths for multi-task training complicates the base model training.

While most previous works focus on more generic cases of domain mismatch, we focus specifically on the domain mismatch between the domain on which the pre-trained models were trained and the domain on which they are fine-tuned in the context of SE. 
SE models utilize pairs of clean and noisy speech for training. Under the assumption that pre-trained speech models are trained using clean data, the paired fine-tuning data can be treated as a pair consisting of source-domain data (clean speech) and target-domain data (noisy speech). Thus, during fine-tuning, we can extract both source- and target-domain statistics. Our proposed method is related to utilizing these estimated statistics without requiring any additional trainable variables or pre-training phases. 
Our method shifts the distribution of the latent features from the target-domain to the source-domain to alleviate domain conflicts during fine-tuning. To smoothly adjust the model from source-domain features to target-domain features, we decrease the shifting factor during the fine-tuning phase.

\section{Base model}
\label{sec:base}
For our experiments, we used several Mockingjay variants (Mockingjay, TERA) and wav2vec 2.0 variants (wav2vec 2.0, WavLM, HuBERT) as the base upstream models. The Mockingjay variants use generative learning methods and the wav2vec 2.0 variants use contrastive learning methods.
We followed the BASE setups for all of the models, and implemented appropriate SE networks depending on their architectures and the structure of the projection layers.

\subsection{Base Mockingjay downstream}
\label{sec:mockingjayvar}
Mockingjay is a speech representation network which consists of a multi-layer transformer encoder \cite{liu2020mockingjay}. During the pre-training phase, the output of the transformer encoder is fed to feed-forward projection layers to predict masked frames.
Given this, we implemented an SE network by adding a single convolutional layer on top of the body, which is trained to generate the real and imaginary parts of cIRM masks \cite{williamson2015complex}.
In other words, our downstream model is trained to predict the complex masks of each time-frequency bin, and the estimated complex masks are multiplied with the complex noisy inputs to obtain noise suppressed outputs. 

\subsection{Base wav2vec 2.0 downstream}
\label{sec:wav2vec2var}
Wav2vec 2.0 consists of a multi-layer feature encoder followed by a transformer encoder \cite{baevski2020wav2vec}. In the pre-training phase, latent features extracted from the feature encoder are quantized, and they are used as a target for training with a BERT-style objective \cite{kenton2019bert}. 
To implement the downstream SE model, we added a U-Net style decoder \cite{stoller2018wave} that would utilize representations from the transformer encoder and the feature encoder.
Specifically, we used the output of the first layer of the transformer encoder as the bottleneck feature of our SE model. This took into account prior work that found lower layers to contribute more informative features for the SE task \cite{huang2022investigating}.
From the bottleneck feature, a stack of deconvolution layers are paired with the output of the feature encoder layers in reverse order. The features from the feature encoders are passed into point-wise convolutions for dimension reduction, concatenated with the corresponding decoder features and passed to the next deconvolution layers.
\section{Feature normalization}
\label{sec:featnorm}

\begin{table*}[]
  \caption{Performance of various models. For generative models (Mockingjay, TERA), the first layer is normalized. Second layer is normalized in case of wav2vec 2.0 variants. \label{table:1}}
  \centering
  \begin{tabular}{c||c|c|c|c|c}
  \toprule
    Methods & \multicolumn{1}{c|}{WB-PESQ} & \multicolumn{1}{c|}{STOI} & \multicolumn{1}{c|}{CSIG} & \multicolumn{1}{c|}{CBAK} & \multicolumn{1}{c}{COVL} \\ 
  \midrule\midrule
  Mockingjay(TERA)-base & 2.3872 & 0.9077 & 3.7864 & 2.9504 & 3.0675 \\
  \midrule
  Mockingjay-pretrained & 2.9349 & \textbf{0.9440} & 4.3145 & 2.7592 & 3.6241 \\
  Mockingjay-normed & \textbf{3.0096} & 0.9401 & \textbf{4.3500} & \textbf{3.4208} & \textbf{3.6810} \\
  \midrule
  TERA-pretrained & 2.9909 & \textbf{0.9439} & 4.3539 & 2.6352 & 3.6738 \\
  TERA-normed & \textbf{3.0729} & 0.9438 & \textbf{4.3915} & \textbf{2.6941} & \textbf{3.7336} \\
  \midrule\midrule
  wav2vec2.0-base & 2.6646 & 0.9399 & 4.1113 & 2.9734 & 3.3905 \\
  \midrule
  wav2vec2.0-pretrained & 2.7223 & 0.9394 & 4.1793 & 3.0136 & 3.4556 \\
  wav2vec2.0-normed & \textbf{2.7861} & \textbf{0.9437} & \textbf{4.2531} & \textbf{3.0497} & \textbf{3.5290} \\
  \midrule
  wavlm-pretrained & 2.6687 & 0.9418 & 4.1187 & 3.2075 & 3.3973 \\
  wavlm-normed & \textbf{2.7796} & \textbf{0.9440} & \textbf{4.2485} & \textbf{3.2924} & \textbf{3.5222} \\
  \midrule
  HuBERT-pretrained & 2.7106 & 0.9436 & 4.1905 & 3.2420 & 3.4572 \\
  HuBERT-normed & \textbf{2.7838} & \textbf{0.9447} & \textbf{4.2474} & \textbf{3.2513} & \textbf{3.5241} \\
  \midrule\midrule
  DEMUCS \cite{defossez2020real} & 3.0700 & 0.9500 & 4.3100 & 3.4000 & 3.6300 \\
  SE-Conformer \cite{kim2021se} & 3.1300 & 0.9500 & 4.4500 & 3.5500 & 3.8200 \\
  \bottomrule
  \end{tabular}
\end{table*}

\begin{figure}[t]
  \centering
  \includegraphics[width=\linewidth]{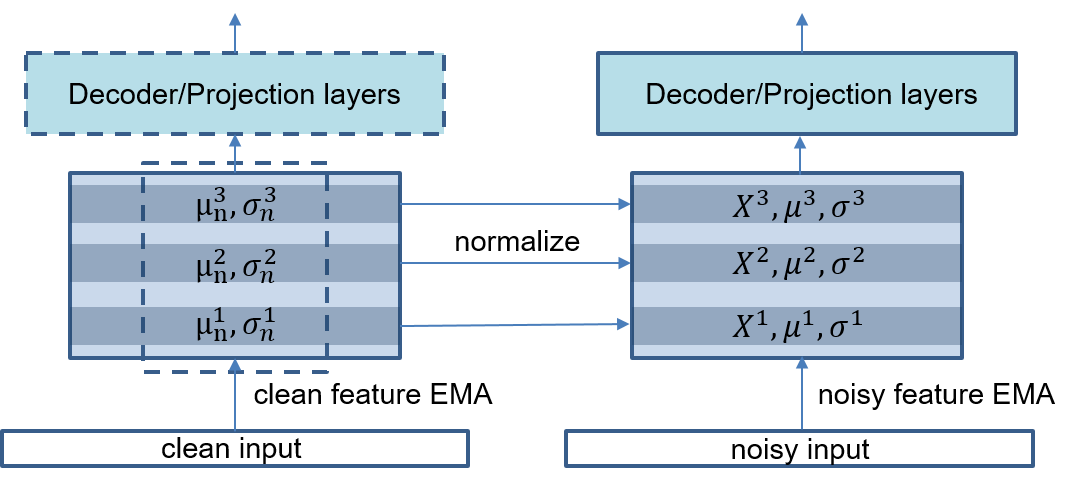}
  \caption{Visualization of the normalization of latent features with accumulated statistics across the training corpus.}
  \label{fig:2}
  \vspace{-8pt}
\end{figure}

Consider a latent feature $X\in\mathcal{R}^d$ from the upstream body of the network under dimension $d$.
Given its mean $\mu\in\mathcal{R}^d$ and standard deviation (std) $\sigma\in\mathcal{R}^d$, we can
calculate the normalized feature $X_n$ with new target mean $\mu_n$ and std $\sigma_n$ as follows:
\begin{align} 
X_n& = \frac{X - {\mu}}{\sigma}{\sigma}_n + {\mu}_n \\
   & = \frac{{\sigma}_n}{\sigma}X + ({\mu}_n - \frac{{\sigma}_n}{\sigma}{\mu}) \\
   & = {r_n}X + ({\mu}_n - r_n{\mu}), \text{ where }{r_n} = {\sigma}_n/{\sigma}.
\end{align}

Fig. \ref{fig:2} shows the proposed feature normalization method that maintains latent statistics $r_n$, ${\mu}_n$ and ${\mu}$ required when normalizing the latent feature. Through normalization, the (noisy) input features follow the same statistics as the corresponding clean features. 
We used the exponentially moving average (EMA) method for each feature to recursively estimate the statistics of the features in the training corpus.

We gradually decrease the normalization effect by controlling a scale factor $k$. $k$ becomes zero when the model is close to reaching convergence, which means that normalization is barely done at this stage. 
This factor scheduling method means that the model does not require additional statistical parameters $r_n$, ${\mu}_n$ and ${\mu}$ during evaluation. The re-normalized feature $\hat{X}_n$ with the scale factor can be calculated as follows:
\begin{align}
  \delta_n & = k(X_n - X) \\
           & = k \left( \frac{{\sigma}_n}{\sigma}-1 \right ) X + k \left ({\mu}_n - \frac{{\sigma}_n}{\sigma}{\mu} \right), \\
  \hat{X}_n& = X + \delta_n \\
           & = \frac{k{\sigma}_n+(1-k){\sigma}}{\sigma} X + k \left ({\mu}_n - \frac{{\sigma}_n}{\sigma}{\mu} \right ).
\end{align}
%Intuitively, $\hat{X}_n$ has a smoothed standard deviation with smoothing weight $k$, and a smoothed mean that is defined by the difference between the reference mean and re-scaled mean. 
The detailed process can be found in Algorithm \ref{alg:1}.

\begin{algorithm}[t]
  \caption{single step of feature normalization}
  \label{alg:1}
  \KwIn{pair of noisy input $x$ and clean output $y$, factor $k$, momentums $beta_m$ and $beta_r$}
  $x_0, y_0 = x, y$\;
  \For{\textup{each layer }$l$\textup{ in upstream body}}{
    \If{\textup{normalize layer }$l$}{
      extract statistics $\hat{\mu}, \hat{\sigma}$ from $x_l$\;
      extract statistics $\hat{\mu_n}, \hat{\sigma_n}$ from $y_l$\;
      $\hat{r} = \hat{\sigma_{n}} / \hat{\sigma}$ \;
      $\mu^l \leftarrow beta^l_m * \mu^l + (1 - beta^l_m) * \hat{\mu}$ \;
      $\mu^l_n \leftarrow beta^l_m * \mu^l_n + (1 - beta^l_m) * \hat{\mu_n}$ \;
      $r^l \leftarrow beta^l_r * r^l + (1 - beta^l_r) * \hat{r}$ \;
      $\hat{x_l} = x_l + k(r^l * x_l + (\mu^l_n - r^l\mu^l) - x_l)$\;
      $x_{l+1} = layer(\hat{x_l})$\;
      $y_{l+1} = layer_{frozen}(y_l)$\;
    }
    \Else{
      $x_{l+1} = layer(x_l)$\;
    }
  }
\end{algorithm}

From an implementation point of view, a frozen duplicate of the upstream body, ranging from the first layer to the layer in which the last normalization occurs, must be maintained, as denoted by $layer_{frozen}$ in Algorithm \ref{alg:1}. The frozen layer is used for extracting clean latent features not affected by the training status. The number of additional frozen parameters retained during training is negligible from the fact that the normalization occurs at the very beginning of each layer.
\section{Experimental Setup}
\label{sec:setup}

\subsection{Data}
We used pre-trained Mockingjay, TERA, wav2vec 2.0, WavLM, and HuBERT trained using the Librispeech-960hr dataset \cite{panayotov2015librispeech}.
For the training and evaluation of the downstream models, we used the Voicebank-DEMAND corpus \cite{valentini2017noisy} after downsampling to 16kHz. Training data was cropped to correspond to 2 seconds or 1 second for the Mockingjay variants and wav2vec 2.0 variants, respectively. The Mockingjay variants were trained for 50k iterations with a batch size of 8, while wav2vec 2.0 variants were trained for 100k iterations with a batch size of 8.
For evaluation, we used WB-PESQ, a wideband version of PESQ \cite{rix2001perceptual}, and STOI \cite{taal2010short}.

\subsection{Model details}
For Mockingjay variants, training was performed with cIRM loss, time-domain MSE loss, and MSTFT loss \cite{yamamoto2020parallel}. 
For cIRM loss, $K = 10$ and $C = 0.1$ are used. 
%as in the original paper. 
For MSTFT loss, we set three sub-losses $L^{(1)}_s$, $L^{(2)}_s$ and $L^{(3)}_s$ with FFT size, window size, and frame shift (1024, 400, 80), (2048, 800, 160), and (512, 160, 32), respectively. 
%The relative weights for each loss were set to 1.
For wav2vec 2.0 variants, training was performed with time-domain MSE loss and MSTFT loss following the same configurations as above.

We linearly decreased the scaling factor $k$ towards zero during training. $k$ was initially set to 0.5 for the wav2vec 2.0 variants, 0.8 for TERA, and 1.0 for Mockingjay. Experiments on different factor scheduling schemes besides linear decay showed no significant differences in output performance.
Momentum for $\mu$ and $r$ were set to 0.99 and 0.999 respectively. 

\begin{table*}[]
  \caption{Performance of TERA and wav2vec 2.0 based downstream models with different normalization layers. \label{table:2}}
  \centering
  \begin{tabular}{c||c|c|c|c|c}
  \toprule
    Methods & \multicolumn{1}{c|}{WB-PESQ} & \multicolumn{1}{c|}{STOI} & \multicolumn{1}{c|}{CSIG} & \multicolumn{1}{c|}{CBAK} & \multicolumn{1}{c}{COVL} \\ 
  \midrule\midrule
  TERA-pretrained                                   & 2.9909 & 0.9439 & 4.3539 & 2.6352 & 3.6738 \\
  TERA-normed (layer 1)          & \textbf{3.0729} & 0.9438 & \textbf{4.3915} & 2.6941 & \textbf{3.7336} \\
  TERA-normed (layer 2)                             & 3.0318 & \textbf{0.9440} & 4.3717 & \textbf{2.7316} & 3.7031 \\
  TERA-normed (layer 3)                             & 3.0028 & 0.9415 & 4.3587 & 2.6455 & 3.6825 \\
  %\setrow{\bfseries} TERA-normed (layer 1\&2)       & 3.0740 & 0.9430 \\
  \midrule
  wav2vec2.0-pretrained                             & 2.7223 & 0.9394 & 4.1793 & 3.0136 & 3.4556 \\
  wav2vec2.0-normed (layer 1)                       & 2.7692 & \textbf{0.9441} & 4.2330 & 2.5209 & 3.5099 \\
  wav2vec2.0-normed (layer 2)    & \textbf{2.7861} & 0.9437 & \textbf{4.2531} & \textbf{3.0497} & \textbf{3.5290} \\
  wav2vec2.0-normed (layer 3)                       & 2.3002 & 0.9219 & 3.7291 & 2.8974 & 3.0007 \\
  wav2vec2.0-normed (layer 4)                       & 2.6417 & 0.9376 & 4.1099 & 2.9611 & 3.3756 \\
  %*\setrow{\bfseries} wav2vec2.0-normed (layer 1\&2) & &  \\
  \bottomrule
  \end{tabular}
%  \vspace{-8pt}
\end{table*}
\section{Results}
\label{sec:results}

\subsection{Results on downstream models}
Table \ref{table:1} shows the results of applying our proposed feature normalization to several upstream models. Methods with the suffix ``base" refer to a model with random weight initialization. Models are tagged as ``pretrained" if they are initialized with pre-trained weights, and further noted as ``normed" if feature normalization is applied during training. The results show that feature normalization improves both WB-PESQ and STOI for almost every type of model, demonstrating that the feature normalization scheme is effective and model agnostic.
We also note that for the generative models, performance nearly matches that of other SE-specific architectures including DEMUCS \cite{defossez2020real} and SE-Conformer \cite{kim2021se}.

One interesting aspect of the results is the improvement ratios between the generative and contrastive models. For generative models, initial baseline performance tends to be lower than that of the contrastive models, but their performance improves significantly after loading pre-trained weights. This is possibly due to the generative models being trained to fill in masked frames, which causes them to capture more local information between frames, whereas contrastive models tend to learn high-level semantics which capture global-level dependencies.

\subsection{Effect of normalization layer}
Table \ref{table:2} shows the results when normalizing each layer of TERA and wav2vec 2.0. We can see that the improvement rate increases when normalization is applied to the lower layers. 
This is perhaps unsurprising, given that domain mismatch has the greatest impact on performance at lower layers. Additionally, we hypothesize that the non-linearities introduced by stacking multiple layers with non-linear activations contributes to these results. Instead of linear normalization, non-linear smoothing that takes into account the non-linearities in the overall network may help improve results in later layers. 

\begin{figure}[h]
  \centering
  \includegraphics[width=\linewidth]{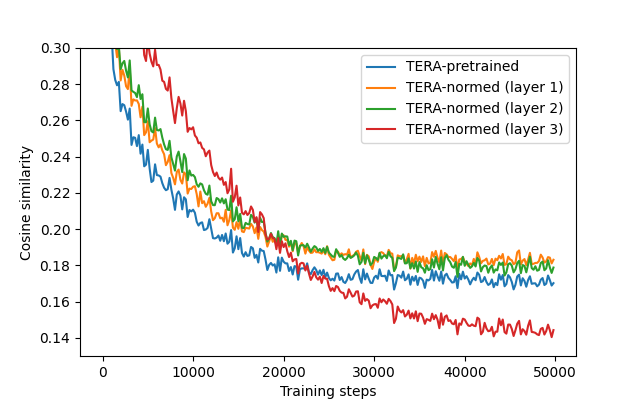}
  \caption{Cosine similarities between the clean feature from and the noisy features with different layer normalizations.}
  \label{fig:cos}
  \vspace{-8pt}
\end{figure}

To further inspect the effect of normalization in each layer, we evaluated cosine similarities of the encoder output features between clean and noisy inputs. Figure \ref{fig:cos} shows the similarities between the clean reference features and the features from the noisy input with normalization applied to each of the layers over the course of training. The features have higher similarities when lower layers are normalized during training. This implies that normalization in the lower layers helps the upstream body maintain the clean feature semantics and avoid catastrophic forgetting, which can occur when domain adaptation is done during fine-tuning.
Conversely, applying normalization to the highest layer led to higher similarities in the early stages of training, but a rapid drop later on. This suggests that choosing proper normalization layers can either make the upstream body maintain its good representative power from the pre-training phase or confuse the network and cause it to lose its ability to generalize.
\section{Conclusions}
\label{ssec:conclusions}

We introduced a feature normalization method for representations from pre-trained speech models that facilitate the use of these representations for downstream speech enhancement. By normalizing the features from noisy inputs to have the same statistics as clean reference inputs, we were able to significantly improve enhancement performance when fine-tuning several different pre-trained speech models. Our results showed that meaningful improvements occur the most when feature normalization is applied only to the lower layers of the pre-trained model. This implies that, for features from higher layers, more sophisticated normalization methods may need to be designed to deal with the various non-linearities in the network. 

For future work, we will study more generic feature normalization that can be applied to representations from multiple layers within a network. Additionally, domain mismatch can arise in speech separation tasks. Speech separation model needs to be trained on input with multiple speakers, but a pre-trained upstream model may have been trained on single-speaker data. To further explore the use of pre-trained models in downstream tasks, we aim to extend our feature normalization approach to speech separation and gain additional insights.

\bibliographystyle{IEEEtran}
\bibliography{refs}

\end{document}